\documentclass[showpacs,amsmath,pra]{revtex4}
\usepackage{graphicx}
\usepackage{dcolumn}
\usepackage{bm}

\begin{document}

\title{Quantum dialogue}
\author{Nguyen Ba An}
\email{nbaan@kias.re.kr}
\affiliation{School of Computational Sciences, Korea Institute for Advanced Study,\\
 207-43 Cheongryangni 2-dong, Dongdaemun-gu, Seoul 130-722, Republic of  Korea }

\begin{abstract}
We propose an entanglement-based protocol for two people to simultaneously exchange 
their messages. We show that the protocol is asymptotically secure against the disturbance 
attack, the intercept-and-resend attack and the entangle-and-measure attack. Our protocol 
is experimentally feasible within current technologies.
\end{abstract}

\maketitle

Sending or/and exchanging secret information 
has long been desired since language became a tool to communicate. 
Up to date the most popular cryptosystem is the RSA protocol 
\cite{rsa} whose security is based upon unproven mathematical assumptions,
e.g., it is extremely hard to factorize a large integer. Because such
a mathematically difficult task could be accomplished by an efficient quantum
computation algorithm \cite{shor}, all the RSA-based privacy would be broken if
scalable quantum computers come into being some day. Fortunately enough, however, 
laws of quantum mechanics can also be exploited to make provably secure distribution
of secret information. This is known as quantum cryptography. Conventionally
the problem reduces to the so called quantum key distribution (see, e.g., 
\cite{qkd}) which is \textit{nondeterministic} since one never knows which
transmitted bits will actually be used and which should be discarded during the
distribution, and the number of discarded bits is at least one half
of the total processed bits. Furthermore, the real message can only be
read after the secret key (i.e., a sequence of random bits whose
length is equal to that of the message) is established and shared between the two
legitimate parties.

Recently, quite different quantum crytographic scenarios have been proposed \cite
{s1,s2,s3,s4,s5,s6} for secure communication \textit{without} a prior secret key
distribution. In particular, the so called ping-pong protocol (PPP) \cite{s3}
allows the encoded bit to be decoded \textit{instantaneously} 
in each respective transmission run. In other words, the PPP provides 
a quantum means of \textit{direct} and \textit{deterministic} communication. 
Nevertheless, the PPP supports only one-way communication and contains 
in itself some limitation. 

In this letter we first point out a drawback of the original PPP and then improve it 
towards a protocol, called quantum dialogue protocol, which enables both 
legitimate parties (Alice and Bob) to exchange 
their secret messages in a direct way, much like in a dialogue. 

In the original PPP \cite{s3} Bob is provided with a number of Einstein-Podolsky-Rosen (EPR) 
pairs \cite{ein}, all in the entangled state  
\begin{equation}
\left| \Psi_{0,0}\right\rangle _{ht}=\frac{1}{\sqrt{2}}\left( \left|
\downarrow \right\rangle _{h}\left| \uparrow \right\rangle
_{t}+\left| \uparrow \right\rangle _{h}\left| \downarrow
\right\rangle _{t}\right),\label{EPR}
\end{equation}
where $h$ stands for ``\textit{home}'', $t$ for ``\textit{travel}''
while $\left| \downarrow \right\rangle$ and $\left| \uparrow \right\rangle $
characterize two degrees of freedom of a qubit. 
In each run Bob keeps qubit $h$ and ``pings" qubit $t$ to Alice. Alice encodes her information 
by performing $C^t_{0,0}=\hat{1}^t$ or $C^t_{1,1}=\sigma^t_z$ 
($\sigma^t_{x,y,z}$ the Pauli matrices) on the qubit $t$ depending 
on the value of her message bit is ``0" or ``1", then ``pongs" the qubit $t$ back to Bob who is able 
to decode Alice's secret bit with certainty by a Bell measurement on the $ht$-pair. 
This is a message mode (MM). To check for eavesdropping Alice and Bob 
sometimes agree to switch to 
a control mode (CM) in which Alice measures qubit $t$ in the bases  
$\mathcal{B}=\{\left| \downarrow \right\rangle ,\left| \uparrow \right\rangle \}$, then, 
instead of ``ponging", publicly announces her measurement outcome to Bob who 
can probabilistically detect the presence of Eve (the eavesdropper) by measuring 
qubit $h$ (also in the bases $\mathcal{B}$) and comparing his measurement outcome 
with Alice's.    
  
The serious drawback suffered by the PPP is the following. 
Since MM operates in a ``ping-pong" manner while CM operates just in a ``ping" one, 
Eve can easily avoid all control runs and manipulate qubits $t$ in MM in such 
a way as to totally disturb the message meaning. For that purpose, Eve waits on 
the pong-route. If a qubit $t$ comes out from Alice this is surely a run in MM. 
Eve may simply either measure the qubit $t$ \cite{s5} or 
\textit{randomly} apply either $C^t_{0,0}$ or $C_{1,1}^{t}$ on it. 
In the first situation the entanglement between qubits $h$ and $t$ is destroyed. In the
second situation the phase of the EPR-pair changes \textit{randomly}. In both
situations Eve remains undetected and what Bob decodes is nothing else but a random sequence of bits
that contains no information at all.  This is a kind of denial-of-service attacks. 
Here we refer to it as disturbance attack for short. 

To protect against such a disturbance attack we modify the CM as follows. 
After manipulation of a qubit $t$ Alice \textit{always}  ``pongs" the qubit $t$ back to Bob. 
The modified CM  is a mode in which Alice lets Bob know her encoding 
transformation which then allows Bob to detect Eve 
by analyzing the outcome of his Bell measurement on the EPR-pair. The point of the modification   
is that Eve cannot distinguish between MM and CM since both modes now operate in the same 
``ping-pong" manner (compare with the original PPP in which MM and CM are distinguishable: 
MM is ``ping-pong" like but CM is ``pong" like).   
 
The modified PPP is good against the disturbance attack mentioned above but it is insecure by 
the following intercept-and-resend attack. 
On the ping-route  Eve gets the qubit $t$ and keeps it with her. Afterwards she creates her own 
entangled pair in the same state as in Eq. (\ref{EPR}), i.e., Eve's pair state is 
\begin{equation}
\left| \Psi_{0,0}\right\rangle _{HT}=\frac{1}{\sqrt{2}}\left( \left|
\downarrow \right\rangle _{H}\left| \uparrow \right\rangle
_{T}+\left| \uparrow \right\rangle _{H}\left| \downarrow
\right\rangle _{T}\right) ,
\end{equation}
and sends her qubit $T$ to Alice. Alice would take 
$T$ for $t$, encodes her message bit by performing an appropriate transformation 
as described above and ``pongs" the qubit $T$ back to Bob. 
On the pong-route Eve gets back the transformed qubit $T$, carries out a Bell measurement on the $HT$-pair 
to learn Alice's secret bit.  By the same Bell measurement Eve knows the encoding transformation Alice performed 
on the qubit $T$. Eve then applies the same transformation 
on the qubit $t$ she has kept and ``pongs" it back to Bob. Clearly, Alice's message is 
readable not only to Bob but also to Eve and, even worse, Eve's tampering is absolutely unnoticeable.     

To rescue the modified PPP against the intercept-and-resend attack 
we further improve it in such a way so that the initial 
entangled pairs of Bob are not all in the same state $\left| \Psi_{0,0}\right\rangle _{ht}$ 
but they must be somehow chosen each time as one among the four mutually orthogonal Bell states 
$\left| \Psi_{k,l}\right\rangle _{ht}=C^t_{k,l}\left| \Psi_{0,0}\right\rangle _{ht}$ where 
$C^t_{0,0}$, $C^t_{0,1}$, $C^t_{1,0}$ and $C^t_{1,1}$ denote $\hat{1}^t$, $\sigma_x^t$, 
$\sigma_y^t$ and $\sigma_z^t$, respectively. The choice may be random or in some secret 
fashion unknown to Eve. The latter alternative suggests a quantum dialogue protocol which will be detailed 
now. 

Suppose that Alice has a secret message consisting of $2N$ bits 
\cite{2N},
\begin{equation}
{\rm{Alice's}}\,\,{\rm{message}} =\{(i_{1},j_{1}),(i_{2},j_{2}),\ldots ,(i_{N},j_{N})\},  \label{X}
\end{equation}
with $i_{n},j_{n}\in \{0,1\}$ and, Bob has another secret message consisting of $2M$ bits \cite{2N},
\begin{equation}
{\rm{Bob's}}\,\,{\rm{message}}=\{(k_{1},l_{1}),(k_{2},l_{2}),\ldots ,(k_{M},l_{M})\},  \label{Y}
\end{equation}
with $k_{n},l_{n}\in \{0,1\}$. Without loss of generality we can set $N=M$ \cite{MN}.
To securely exchange their messages or, in other words, to carry out a secret dialogue, 
 Bob first produces a large enough 
number of entangled pairs, all in the state $\left| \Psi_{0,0}\right\rangle _{ht}$, Eq. (\ref{EPR}). 
 Then Bob and Alice proceed as follows.

\begin{description}
\item[1.]  $Set$ $n=0.$

\item[2.]  $Set$ $n=n+1.$ Bob encodes his bits $(k_n,l_n)$ by applying 
$C^t_{k_n,l_n}$ on the state $\left| \Psi_{0,0}\right\rangle _{h_nt_n}$, keeps 
qubit $h_n$  with him and pings qubit  $t_{n}$  to Alice. Then 
Bob lets Alice know that \cite{bob}. 

\item[3.]  Alice confirms Bob that she received a qubit \cite{alice}. 
Then she encodes her bits $(i_n,j_n)$ by performing the transformation 
$C^t_{i_n,j_n}$ on that qubit and pongs it back to Bob.  

\item[4.] Having been aware of Alice's confirmation, 
Bob performs a Bell measurement on the two qubits \cite{HT}  
with the result in state $\left| \Psi_{x_n,y_n}\right\rangle _{h_nt_n}$ 
 $(x_n,y_n \in [0,1])$, and waits for Alice to tell him that was a run in MM or in CM.

\begin{description}
\item[4.1.]  If it was a MM run, Bob decodes Alice's bits as $(i_{n}=|x_n-k_n|, j_{n}=|y_n-l_n|)$,  
then publicly announces the values of  $(x_n,y_n)$ to allow Alice also to decode Bob's bits as 
$(k_{n}=|x_n-i_n|, l_{n}=|y_n-j_n|)$. 
Afterwards the protocol proceeds to Step 5 if $n=N$ or to Step 2 if $n<N$. 

\item[4.2.]  If it was a CM run, Alice publicly reveals the value of $(i_n,j_n)$ 
for Bob to check the eavesdropping: 
if both $i_n=|x_n-k_n|$ and $j_n=|y_n-l_n|$ hold, the process continues, 
i.e., Bob sets $n=n-1$ and goes to Step 2; otherwise 
the process is reinitialized by going to Step 1.
\end{description}

\item[5.] The dialogue has been successfully completed.
 
\end{description}

We now explicitly analyze the quantum dialogue protocol described above. 
After Bob encodes his bits $(k,l)$ on the EPR-pair state $\left| \Psi_{0,0}\right\rangle _{ht}$, 
the pair state becomes $\left| \Psi_{k,l}\right\rangle _{ht}$, i.e.,
\begin{equation} 
\left| \Psi_{0,0}\right\rangle _{ht}\rightarrow 
\left| \Psi_{k,l}\right\rangle _{ht}=C^t_{k,l}\left| \Psi_{0,0}\right\rangle _{ht}. 
\end{equation}
Since
\begin{equation}
C^t_{i,j}C^t_{k,l}=\phi_{i,j;k,l} C^t_{i\oplus k,j\oplus l},
\end{equation}
where the $\oplus$ denotes an addition mod 2 and $\phi_{i,j;k,l}$ is a phase factor 
($\phi_{i,j;k,l} =1$ or $\pm i$ depending on the values of $i,j,k,l$ \cite{alpha}). 
Further, after Alice's encoding,  the state $\left| \Psi_{k,l}\right\rangle _{ht}$ of the qubit pair 
 is transformed as  
 \begin{equation} 
\left| \Psi_{k,l}\right\rangle _{ht}\rightarrow 
C^t_{i,j}\left| \Psi_{k,l}\right\rangle _{ht} =C^t_{i,j}C^t_{k,l}\left| \Psi_{0,0}\right\rangle _{ht}
= \phi_{i,j;k,l}\left| \Psi_{i\oplus k,j\oplus l}\right\rangle _{ht}. 
\end{equation}
Clearly, if the outcome of Bob's Bell measurement is $(x,y)$, then Bob can easily 
decode Alice's bits as $(i=|x-k|, j=|y-l|)$ because Bob knows his bits $(k,l)$.  
At the same time, Alice can also easily 
decode Bob's bits as $(k=|x-i|, l=|y-j|)$ because Alice knows her bits $(i,j)$ 
and the values of $(x,y)$ broadcasted by Bob.  
The crucial merit is that, although Eve knows $(x,y)$ as well (through Bob's public 
broadcasting), she can by no means, except a pure guess, read either 
Alice's or Bob's message because none of the bits $(i,j,k,l)$ are known to her. 
By the same reason Eve faces a detection probability of $3/4$ per CM run in both 
the disturbance attack and the intercept-and-resend attack. Let $\mathcal{N}$
be the total number of the protocol runs among which there are $\mathcal{N}_{MM}$ 
runs in MM and $\mathcal{N}_{CM}$ runs in CM: 
$\mathcal{N}=\mathcal{N}_{MM}+\mathcal{N}_{CM}$. The probability of a CM run 
is thus $c=\mathcal{N}_{CM}/\mathcal{N}$. For $\mathcal{N}=1,2,3,\ldots$ the probability 
of detecting Eve is $3c/4$, $3c/4+3c(1-3c/4)/4$, $3c/4+3c(1-3c/4)/4+3c(1-3c/4)^2/4$, $\ldots$, 
respectively.  Therefore, after $\mathcal{N}$ runs the total detection probability is 
\begin{equation}
D=\frac{3c}{4}\sum_{n=0}^{\mathcal{N}-1}\left(1-\frac{3c}{4}\right)^n=
1-\left(1-\frac{3c}{4}\right)^{\mathcal{N}}.\label{D}
\end{equation}
For messages of $2N$ bits (see Eqs. (\ref{X}) and (\ref{Y})) to be entirely exchanged, Alice 
and Bob need $\mathcal{N}_{MM}=N$ runs in MM.  
Taking this into account we can re-express Eq. (\ref{D}) in terms of $N$ (i.e., of message half-length) as
\begin{equation}
D=1-\left(1-\frac{3c}{4}\right)^{\frac{N}{1-c}}.
\end{equation}
Transparently, for any possible value of $c$ $(0<c<1)$, $D$ tends to unity 
in the limit of large $N$ (long message). The greater the value of $c$ 
the higher the speed at which $D$ approaches unity.    

Besides the two above-discussed attacks, 
there is another kind of attack by which Eve could gain a partial information. 
Let us call it entangle-and-measure attack which acts in the following way. 
Eve prepares an ancilla in the initial 
state $\left| \chi\right>_e$ and waits in the ping-route. After Bob applies 
$C^t_{k,l}$ Eve entangles her ancilla with the qubit $t$ by performing 
an unitary operation $\mathcal{E}^{te}$ defined as
\begin{eqnarray}
\mathcal{E}^{te}\left| \downarrow \right\rangle _{t}\left| \chi \right\rangle _{e}
&=&\alpha \left| \downarrow \right\rangle _{t}\left| \chi
_{0}\right\rangle _{e}+\beta \left| \uparrow \right\rangle _{t}\left| \chi
_{1}\right\rangle _{e}, \\
\mathcal{E}^{te}\left| \uparrow \right\rangle _{t}\left| \chi \right\rangle _{e}
&=&\alpha \left| \uparrow \right\rangle _{t}\left| \chi
_{0}\right\rangle _{e}+\beta \left| \downarrow \right\rangle _{t}\left| \chi
_{1}\right\rangle _{e},
\end{eqnarray}
with $\alpha ,\beta $ (assumed to be real) satisfying the normalization
condition $\alpha ^{2}+\beta ^{2}=1$ and  $\{\left| \chi _{0}\right\rangle
_{e}, $ $\left| \chi _{1}\right\rangle _{e}\}$ being the pure orthonormalized 
ancilla's states uniquely determined by the unitary operation $\mathcal{E}^{te}$. 
Subsequently, Eve lets the qubit $t$ going on to Alice. The total system state 
($ht$-pair plus ancilla) before reaching Alice is
\begin{eqnarray}
\left|\Phi_{ping}\right>_{hte}
&=&\mathcal{E}^{te}C^t_{k,l}\left|\Psi_{0,0}\right>_{ht}\left|\chi\right>_e
\nonumber \\
&=&\alpha \left|\Psi_{k,l}\right>_{ht}\left|\chi_0\right>_e
+\beta\phi_{0,1;k,l}\left|\Psi_{k,1\oplus l}\right>_{ht}\left|\chi_1\right>_e.
\end{eqnarray}
On the pong-route, after Alice  encodes her bits by $C^t_{i,j}$ Eve measures her ancilla 
in attempt to gain Alice's information. Since the total system state at the measurement time 
is given by  
\begin{eqnarray}
\left|\Phi_{pong}\right>_{hte}&=&C^t_{i,j}\left|\Phi_{ping}\right>_{hte}
\nonumber \\
&=&\alpha \phi_{i,j;k,l}\left|\Psi_{i\oplus k,j\oplus l}\right>_{ht}\left|\chi_0\right>_e
+\beta\phi_{0,1;k,l}\phi_{i,j;k,1\oplus l}\left|\Psi_{i\oplus k,j\oplus 1\oplus l}\right>_{ht}
\left|\chi_1\right>_e,\label{pong}
\end{eqnarray}
Eve conceals herself if her measurement outcome ends up with $\left|\chi_0\right>_e$ 
(with probability $\alpha^2$). However, if she finds $\left|\chi_1\right>_e$ 
(with probability $\beta^2$), then $\left|\Phi_{pong}\right>_{hte}$ collapses into 
$\left|\Psi_{i\oplus k,j\oplus 1\oplus l}\right>_{ht}$ which is 
orthogonal to $\left|\Psi_{i\oplus k,j\oplus l}\right>_{ht}$. 
Obviously, this enables Bob to detect Eve in a CM run. It is also clear that the detection probability 
of the entangle-and-measure  attack is $\beta^2$ per CM run and $c\beta^2$ per protocol run. 
Therefore the proposed dialogue protocol is also asymptotically secure against the entangle-and-measure 
attack since its total detection probability $D'$, 
\begin{equation}
D'=1-(1-c\beta^2)^{\frac{N}{1-c}},
\end{equation}
approaches unity in the long-message limit for any possible values of $c$ and $\beta $.

To evaluate how much information Eve could gain when there is no control run 
we calculate the \textit{von Neumann} entropy 
$S(\rho_e)$ of the Eve's reduced density matrix $\rho_e$. From Eq. (\ref{pong}), we obtain
\begin{equation}
\rho_e={\rm{Tr}}_{ht}\left( \left|\Phi_{pong}\right>_{hte}\left<\Phi_{pong}\right|\right)
=\alpha^2\left|\chi_0\right>_e\left<\chi_0\right|+\beta^2\left|\chi_1\right>_e\left<\chi_1\right| .
\end{equation} 
Hence
\begin{equation}
S(\rho_e)=-(1-\beta^2)\log_2(1-\beta^2)-\beta^2\log_2\beta^2. \label{S}
\end{equation}
It follows from Eq. (\ref{S}) that $S>0$ iff $\beta^2>0$ ($\beta^2 \in [0,0.5]$), i.e., iff $D'>0$. 
This means that any attempt to steal information causes a non-zero detection probability. 
To be undetected Eve should set $\beta =0$. But by doing so  no entanglement exists 
between the qubit $t$ and the ancilla and, as a consequence, absolutely no information 
is leaked to Eve.

In conclusion, we have proposed a quantum protocol for two legitimate parties to simultaneously 
exchange their secret messages. It modifies and improves the existing ping-pong protocol \cite{s3} 
making subtle use of the superdense coding \cite{sdc} to double the quantum channel capacity.  
In fact, by the wise manipulation of two public bits $(x,y)$ combined with a single-qubit $t$ in a 
state of two entangled qubits, in a MM run four secret bits $(i,j,k,l)$ can be processed: 
each party is able at the same time to send two secret bits as well as to read two other secret bits. 
Our protocol is shown to be asymptotically secure 
(i.e., the detection probability tends to $1$ in the long-message limit) 
against the disturbance attack, the intercept-and-resend attack as well as the  
entangle-and-measure attack. As in the ping-pong protocol, our protocol is deterministic in the sense that, 
in the course of running, the participating parties surely know which run is in MM and which run is 
in CM. While qubits in CM are deterministically discarded, qubits in MM are read directly by both parties 
without a prior quantum key sharing. This looks like that Alice and Bob are ``talking" to each other bit by bit, 
much as though a dialogue is going on between them; so comes the name ``quantum dialogue". 
In contrast to the ping-pong protocol here, 
thanks to the modification of CM, there is no need for both Alice and Bob to do single-qubit measurements. 
Hence, our quantum dialogue protocol  seems even more feasible within present technologies as compared 
to the ping-pong protocol (see, e.g.,  the experimental feasibility of the ping-pong protocol in \cite{s3}).

The author thanks C. H. Bennett, H. W. Lee and J. Kim for useful discussions. 
Support from the KIAS Quantum Information Group is also gratefully acknowledged. 
This research is funded by the KIAS R\&D grant No. 03-0149-002.


\begin{thebibliography}{99}
\bibitem{2N}  For a message of an odd number of bits one can add one
auxiliary bit at the end of the message to make the total bit number even. The value 
of the auxiliary bit is not of any interest at all.
\bibitem{MN} If $N>M$ one can add $2(N-M)$ auxiliary bits $(k_{M+1},l_{M+1})$, 
$(k_{M+2},l_{M+2})$, $\ldots $, $(k_{N},l_{N})$ to Bob's message. 
 If $M>N$ one can add $2(M-N)$ auxiliary bits $(i_{N+1},j_{N+1})$, 
$(i_{N+2},j_{N+2})$, $\ldots $, $(i_{M},j_{M})$ to Alice's message.  
The values of the auxiliary bits are not of any interest at all.
\bibitem{bob} Letting Alice know that Bob has sent his qubit is necessary to avoid 
the unwanted circumstance under which Eve pretends to be Bob all the time without Alice's 
recognization. If such circumstance happens, Eve can without errors read Alice's 
message while Alice just gets Eve's message but not Bob's message since Bob stands 
outside the game.
\bibitem{alice}Alice's confirmation is necessary to avoid 
the unwanted circumstance under which Eve pretends to be Alice all the time without Bob's 
recognization. If such circumstance happens, Eve can without errors read Bob's 
message while Bob just gets Eve's message but not Alice's message since Alice stands 
outside the game. Note also that the qubit Alice receives may come either from Bob or from 
Eve via her intercept-and-resend attack in the ping-route. However, there is no problem as 
Eve's such attack is to be detected by the protocol CM.
\bibitem{HT} Among the two qubits on which Bob performs the Bell measurement 
one has been with him and the other may come either from Alice  or from 
Eve via her intercept-and-resend attack in the pong-route. However, there is no problem as 
Eve's such attack is to be detected by the protocol CM.
\bibitem{alpha}  $\phi_{i,j;k,l}=1$ if $(i=j=0)$ or $(k=l=0)$ 
or $(i=j=k=l=0)$ or $(i=k,j=l)$, $\phi_{i,j;k,l}=i$ if $(i,j)$ and $(k,l)$ 
follows a cyclic sequence $(0,1)\rightarrow (1,0)\rightarrow (1,1)\rightarrow (0,1)$ 
and,  $\phi_{i,j;k,l}=-i$ if $(i,j)$ and $(k,l)$ 
follows a reverse cyclic sequence $(0,1)\rightarrow (1,1)\rightarrow (1,0)\rightarrow (0,1)$.
\end{thebibliography}
\end{document}